# Spectral Shaping with Integrated Self-Coupled Sagnac Loop Reflectors

David J. Moss

*Abstract*— We propose and theoretically investigate integrated photonic filters based on coupled Sagnac loop reflectors (SLRs) formed by a self-coupled wire waveguide. By tailoring coherent mode interference in the device, three different filter functions are achieved, including Fano-like resonances, wavelength interleaving, and varied resonance mode splitting. For each function, the impact of device structural parameters is analyzed to facilitate optimized performance. Our results theoretically verify the proposed device as a compact multi-functional integrated photonic filter for flexible spectral shaping.

## I. INTRODUCTION

WITH a compact footprint, flexible topology, and high scalability, integrated photonic resonators (IPRs) have enabled diverse functional optical devices such as filters, modulators, sensors, switches, and logic gates [1, 2]. As compared with IPRs based on subwavelength gratings [3] and photonic crystal structures [4] that have submicron cavity lengths, IPRs formed by directional-coupled wire waveguides with longer cavity lengths (typically > 10 μm) have smaller free spectral ranges (FSRs) that match with the spectral grids of the state-of-the-art wavelength division multiplexing (WDM) optical communication systems, thus rendering them more widely applicable to these systems. Moreover, the directional-coupled wire waveguides with longer coupling regions and simpler designs also yield a higher tolerance to fabrication imperfections.

Generally, there are two types of basic building blocks for IPRs formed by directional-coupled wire waveguides. The first is a ring resonator, and the second is a Sagnac loop reflector (SLR). In contrast to ring resonators that involve only unidirectional light propagation, the SLRs allow bidirectional light propagation as well as mutual coupling between the light propagating in opposite directions, thus yielding a more versatile coherent mode interference and spectral response. In addition, a standing-wave (SW) resonator formed by cascaded SLRs has a cavity length almost half that of a travelling-wave (TW) resonator based on a ring resonator with the same FSR, which allows for a more compact device footprint.

In our previous work, we investigated integrated photonic filters based on cascaded SLRs [5, 6] and coupled SLRs [7, 8]. Here, we advance this field by introducing the novel approach of using coupled SLRs formed by a self-coupled wire waveguide. This allows us to achieve versatile spectral responses with a simpler design and a higher fabrication tolerance. We tailor the coherent mode interference to achieve three different filter functions, including Fano-like resonances, wavelength interleaving, and varied resonance mode splitting. The requirements for practical applications are considered in our design. Excellent performance parameters are achieved for each filter function, analysis of the impact of the structural parameters and fabrication tolerance is also provided.

## II. DEVICE STRUCTURE

Fig. 1 illustrates a schematic configuration of the proposed structure, consisting of three SLRs formed by a single self-coupled wire waveguide. The device structural parameters are defined in Table I. To simplify the discussion, we assume that $L_{SLR1} = L_{SLR2} = L_{SLR3} = L_{SLR}$ and $L_1 = L_2 = L$. The resonator is equivalent to three cascaded SLRs (which is an infinite-impulse-response (IIR) filter) when $t_2 = 1$ and a SLR with an interferometric coupler [9] (which is a finite-impulse-response (FIR) filter) when $t_1 = t_3 = 1$. When $t_i$ ($i = 1–3$) $\neq 1$, this device is a hybrid filter consisting of both IIR and FIR filter elements, which allows for versatile coherent mode interference and ultimately a diverse range of spectral responses.

We use the scattering matrix method [5, 7] to calculate the spectral response of the device.- In our calculation, we assume a waveguide group index of $n_g = 4.3350$ for transverse electric (TE) mode and a propagation loss of $α = 55$ m$^{-1}$ (i.e., 2.4 dB/cm) based on our previously fabricated silicon-on-insulator (SOI) devices [5, 6]. The device is designed based on, but not restricted to, the SOI platform.

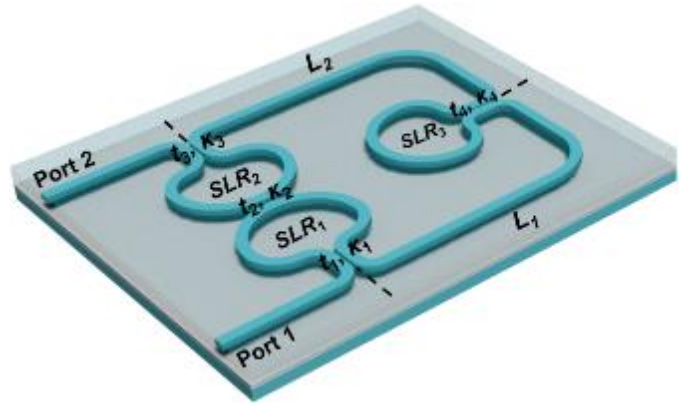

Fig. 1. Schematic configuration of the device configuration consisting of three SLRs ($SLR_1$, $SLR_2$, and $SLR_3$). The definitions of $t_i$ ($i = 1–4$), $k_i$ ($i = 1–4$), $L_{SLRi}$ ($i = 1–3$) and $L_i$ ($i = 1, 2$) are provided in Table I.

D.Moss is with the Optical Sciences Centre, Swinburne University of Technology, Hawthorn, VIC 3122, Australia

TABLE I
DEFINITIONS OF DEVICE STRUCTURAL PARAMETERS

| Waveguides | Length | Transmission factor [a] | Phase shift [b] |
|---|---|---|---|
| Connecting waveguides between SLRs ($i$ = 1, 2) | $L_i$ | $a_i$ | $\varphi_i$ |
| Sagnac loop in $SLR_i$ ($i$ = 1, 2, 3) | $L_{SLRi}$ | $a_{si}$ | $\varphi_{si}$ |
| Directional couplers | | Field transmission coefficient [c] | Field cross-coupling coefficient [c] |
| Coupler in $SLRs$ ($i$ = 1, 3, 4) | | $t_i$ | $\kappa_i$ |
| Coupler between $SLR_1$ and $SLR_2$ | | $t_2$ | $\kappa_2$ |

[a] $a_i$ = exp(-$\alpha L_i$ / 2), $a_{si}$ = exp(-$\alpha L_{SLRi}$ / 2), $\alpha$ is the power propagation loss factor.
[b] $\varphi_i$ = $2\pi n_g L_i$ / $\lambda$, $\varphi_{si}$ = $2\pi n_g L_{SLRi}$ / $\lambda$, $n_g$ is the group index and $\lambda$ is the wavelength.
[c] $t_{si}^2 + \kappa_{si}^2 = 1$ and $t_{bi}^2 + \kappa_{bi}^2 = 1$ for lossless coupling are assumed for all the directional couplers.

## III. FANO-LIKE RESONANCES

Fano resonances that feature an asymmetric spectral lineshape are fundamental physical phenomena that have underpinned many applications such as optical switching, data storage, sensing, and topological optics [10-12]. In this section, the spectral response of the device in Fig. 1 is tailored to realize optical analogues of Fano resonances with high slope rates (SRs) and low insertion loss (IL). The power transmission and reflection spectra with input from Port 1 is depicted in Fig. 2(a-i). The device structural parameters are $L_{SLR} = L = 100$ μm, $t_1 = t_3 = 0.82$, $t_2 = 0.92$, and $t_4 = 1$. Clearly, the output from Port 2 shows periodical Fano-like resonances with an asymmetric resonant lineshape in each period. The high uniformity of the filter shape across multiple periods, or channels, is highly desirable for WDM systems. A zoom-in view of Fig. 2(a-i) is shown in Fig. 2(a-ii), together with another curve showing the corresponding result for another device with the same structural parameters except for a different $t_2 = 1$. As can be seen, when $t_2 = 1$, there is no Fano resonance, distinguishing between the device in Fig. 1 and the three cascaded SLRs in Ref. [5]. The Fano resonances in Fig. 2(a-ii) show a high extinction ratio (ER) of 30.2 dB and a high SR (defined as the ratio of the ER to the wavelength difference between the resonance peak and notch) of 747.64 dB/nm. Table II compares the performance of the Fano-like resonances generated by the coupled SLRs in our previous work [7, 8] and the device in Fig. 1. As compared with previous devices, the device reported here has a much lower insertion loss of 1.1 dB, along with a slightly improved SR. We note that a low IL of 1.1 dB is outstanding among the reported Fano-resonance devices on the SOI platform [13, 14], which renders the device here more attractive for practical applications in optical communication systems.

TABLE II
PERFORMANCE COMPARISON OF FANO-LIKE RESONANCES
GENERATED BY DIFFERENT SLR-BASED DEVICES

| Device structure | IL (dB) | ER (dB) | SR (dB/nm) | FSR (GHz) | Ref. |
|---|---|---|---|---|---|
| Two parallel WC-SLRs [a] | 6.3 | 13.9 | 389 | 692.02 | [7] |
| Three zig-zag WC-SLRs [b] | 3.7 | 63.4 | 721.28 | 230.68 | [8] |
| Device in Fig. 1 | 1.1 | 30.2 | 747.64 | 173 | This work |

[a] WC-SLRs: waveguide coupled SLRs.
[b] For comparison, the length of the SLRs ($L_{SLRi}$, $i$ = 1–3) and the connecting waveguide ($L_i$, $i$ = 1–4) is slightly changed from 115 μm in [8] to 100 μm.

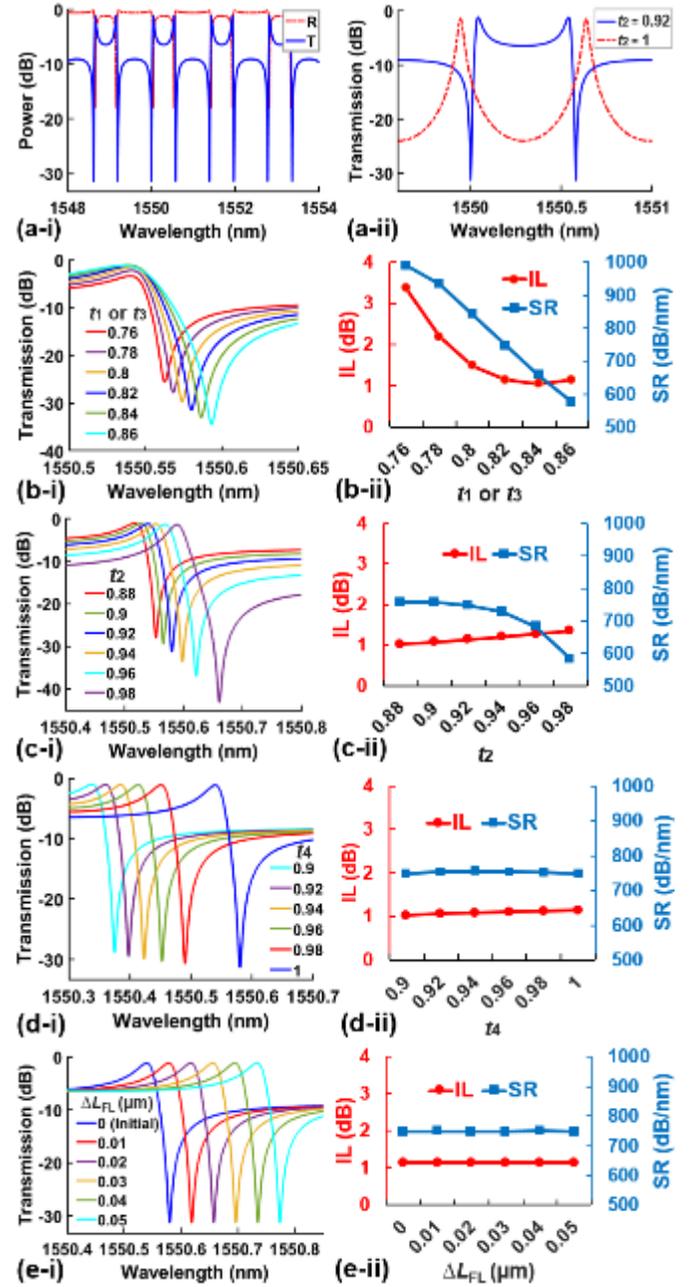

Fig. 2. (a-i) Power transmission and reflection spectra with input from Port 1 when $L_{SLR} = L = 100$ μm, $t_1 = t_3 = 0.82$, $t_2 = 0.92$, and $t_4 = 1$. T: Transmission spectrum at Port 2. R: reflection spectrum at Port 1. (a-ii) Power transmission spectra at Port 2 for $t_2 = 0.92$ and $t_2 = 1$. In (a-ii), the structural parameters are kept the same as those in (a-i) except for $t_2$. (b)–(e) (i) Power transmission spectra and (ii) the corresponding IL and SR for different $t_i$ ($i$ = 1–4) and variation of feedback loop length $\Delta L_{FL}$, respectively. In (b)–(e), the structural parameters are kept the same as those in (a-i) except for the varied parameters.

In Figs. 2(b)–(e), we investigate the impact of the device structural parameters including $t_i$ ($i$ = 1–4) and length variations of the feedback loop ($\Delta L_{FL}$, $L_{FL} = 2L + L_{SLR}$), respectively. In each figure, we changed only one structural parameter, keeping the others the same as those in Fig. 2(a-i). Figs. 2(b-i) and (b-ii) compares the power transmission spectra and corresponding IL and SR for various $t_1$ or $t_3$, respectively. The SR decreases with $t_i$ ($i = 1, 3$), while the IL first decreases with $t_i$ ($i = 1, 3$) and then remains almost unchanged. The spectral response and corresponding IL and SR for different $t_2$ are shown in Figs. 2(c-i) and (c-ii), respectively. The SR decreases with $t_2$, while the IL shows an opposite trend, reflecting that both of the two parameters can be improved by enhancing the coupling strength between $SLR_1$ and $SLR_2$. As shown in Fig. 2(d), both IL and SR remain almost unchanged with varied $t_4$. In Figs. 2(e-i) and (e-ii), we compare the corresponding results for various $\Delta L_{FL}$. As $\Delta L_{FL}$ increases, the filter shape remains unchanged while the resonance redshifts, indicating that the resonance wavelengths can be tuned by introducing thermo-optic micro-heaters [14] or carrier-injection electrodes [15] along the feedback loop to tune the phase shift.

## IV. WAVELENGTH DE-INTERLEAVING

Optical interleavers and de-interleavers are core elements for signal multiplexing and demultiplexing in WDM optical communication systems [16, 17]. In this section, we engineer the spectral response of the device in Fig. 1 to achieve wavelength de-interleaving function. Fig. 3(a) shows the power transmission and reflection spectra with input from Port 1. The device structural parameters are $L_{SLR} = L = 100$ μm, $t_1 = 0.992$, $t_2 = t_3 = 0.95$, and $t_4 = 1$. The input signal is separated into two spectrally interleaved signals, with one group transmitting to Port 2 and the other reflecting back at Port 1. The IL, ER, and 3-dB bandwidth for the passband at Port 2 are 0.36 dB, 12.7 dB, and 83.65 GHz, respectively. The IL, ER, and 3-dB bandwidth for the passband at Port 1 are 0.33 dB, 12 dB, and 91.9 GHz, respectively.

We also investigate the impact of varied $t_i$ ($i = 1$–4), $\Delta L_{FL}$, and $\Delta L_{SLRi}$ ($i = 1, 2$) in Figs. 3(b)–(h), respectively. For simplification, we only show the spectral response at Port 2. In Fig. 3(b), as $t_1$ increases, the ER of the passband decreases while the top flatness improves, reflecting the trade-off between them. In Figs. 3(c)–(e), the bandwidth of the passband increases with $t_2$, $t_3$, $t_4$, respectively, while the ER shows an opposite trend. In Figs. 3(f)–(h), as $\Delta L_{FL}$ or $\Delta L_{SLRi}$ ($i = 1, 2$) increases, the filter shape remains unchanged while the resonance redshifts, indicating the feasibility of achieving tunable de-interleavers with this approach. Since the resonant cavity of the device is formed by a single self-coupled wire waveguide, random length fabrication errors in each part will not induce any asymmetry in the filter shape. This yields a higher fabrication tolerance as compared with the coupled SLRs in Refs. [7, 8], which is particularly attractive for optical interleavers that require a flat-top symmetric filter shape. Note that the de-interleaving function is designed for the telecom C band. According to our previous fabricated devices [17], the slight variation in $t_i$ ($i = 1$–4) arising from the dispersion of silicon would not significantly deteriorate the periodical response across this wavelength range.

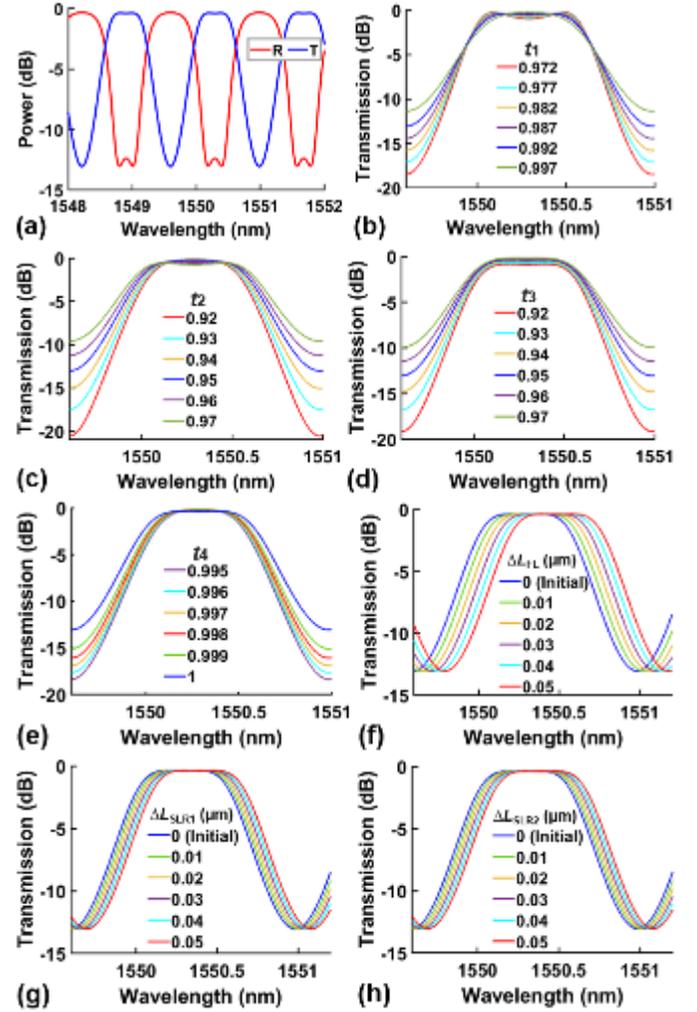

Fig. 3. (a) Power transmission and reflection spectra with input from Port 1 when $L_{SLR} = L = 100$ μm, $t_1 = 0.992$, $t_2 = t_3 = 0.95$, and $t_4 = 1$. T: Transmission spectrum at Port 2. R: reflection spectrum at Port 1. (b)–(h) Power transmission spectra for different $t_i$ ($i = 1$–4), $\Delta L_{FL}$, and $\Delta L_{SLRi}$ ($i = 1, 2$), respectively. In (b)–(h), the structural parameters are kept the same as those in (a) except for the varied parameters.

## V. VARIED RESONANCE MODE SPLITTING

Resonance mode splitting in IPRs induced by coherent mode interference can yield a range of highly useful spectral responses, including electromagnetically induced transparency (EIT), electromagnetically induced absorption (EIA), and Autler–Towns splitting, which have been used for applications such as optical buffering, signal multicasting, analog signal computing, and sensing [9, 18, 19]. In this section, we tailor the spectral response of the device in Fig. 1 to achieve varied resonance mode splitting with diverse spectral response.

Figs. 4(a) and (b) shows the power transmission and reflection spectra for various $t_4$, respectively. The input is from Port 1 and the structural parameters are $L_{SLR} = L = 100$ μm, $t_1 = t_3 = 0.825$, and $t_2 = 0.99$. As can be seen, by increasing the coupling strength of the directional coupler in $SLR_3$ (i.e., reducing $t_4$), the single resonance is gradually split into two resonances with an increased spectral range between them. This is a typical phenomenon for resonance mode splitting similar to those in Ref. [9]. The energy coupling between the light propagating in opposite directions can be changed by varying the reflectivity of SLR$_3$, thus resulting in different mode splitting degrees. Figs. 4(c) and (d) show the power transmission spectrum and group delay response of a Butterworth filter and a Bessel filter formed by resonance mode splitting, respectively. The structural parameters are the same as those in Fig. 4(a) except for a different $t_4$. As shown in Fig. 4(e), the Butterworth filter shape gradually transits to a Chebyshev Type I filter shape by further decreasing $t_1$ or $t_3$. In Fig. 4(f), we compare the spectral response for various $t_2$. It can be seen that more significant resonance mode splitting can be obtained by enhancing the coupling strength between SLR$_1$ and SLR$_2$ (i.e., reducing the $t_2$). In particular, when $t_2 = 1$ (which corresponds to three cascaded SLRs), the resonance is still not split, this indicates that the device reported here shows a significantly enhanced resonance mode splitting as compared with the three cascaded SLRs in Ref. [5].

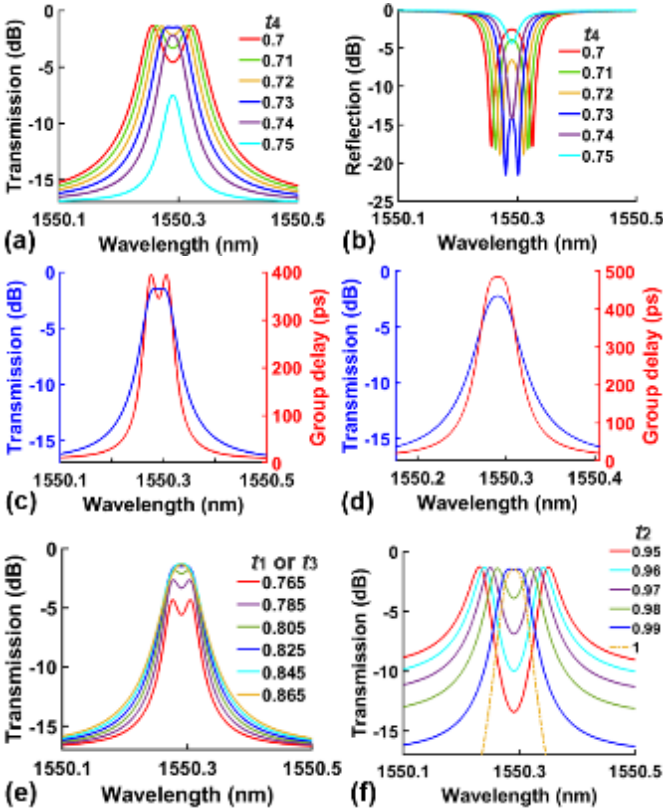

Fig. 4. (a) Power transmission and (b) reflection spectra versus $t_4$ when $t_1 = t_3 = 0.825$, $t_2 = 0.99$, and $L_{SLR} = L = 100$ µm, respectively. The input is from Port 1. (c)–(d) Power transmission spectrum and corresponding group delay response of (c) a Butterworth filter and (d) a Bessel filter when $t_4 = 0.73$ and $t_4 = 0.74$ in (a), respectively. (e)–(f) Power transmission spectra of the Butterworth filter versus $t_1$ or $t_3$, and $t_2$, respectively. In (e)–(f), the structural parameters are kept the same as those in (c) except for the varied parameters.

Finally, this work could have applications to nonlinear devices [20-30] as well as to microwave photonic chips [31-74] and integrated quantum optics [75-89] where advanced optical filter shapes are extremely useful.

## VI. CONCLUSIONS

We theoretically investigate integrated photonic filters based on coupled SLRs formed by a self-coupled wire waveguide. Three different filter functions have been realized, including Fano-like resonances, wavelength interleaving, and varied resonance mode splitting. The compact footprint, versatile spectral responses, and high fabrication tolerance make this approach highly promising for flexible spectral shaping in a diverse range of applications.

Competing interests: The authors declare no competing interests.